\documentclass[%
 aip,
 amsmath,amssymb,
reprint,
]{revtex4-1}

\usepackage{graphicx}
\usepackage{dcolumn}
\usepackage{bm}

\usepackage[utf8]{inputenc}
\usepackage[T1]{fontenc}
\usepackage{mathptmx}
\usepackage{etoolbox}

\makeatletter
\def\@email#1#2{%
 \endgroup
 \patchcmd{\titleblock@produce}
  {\frontmatter@RRAPformat}
  {\frontmatter@RRAPformat{\produce@RRAP{*#1\href{mailto:#2}{#2}}}\frontmatter@RRAPformat}
  {}{}
}%
\makeatother

\usepackage[dvipsnames,table,xcdraw]{xcolor} 
\usepackage[english]{babel}
\usepackage[utf8]{inputenc}
\usepackage[colorinlistoftodos, color=green!40, prependcaption]{todonotes}
\usepackage[section]{placeins}
\usepackage{float}
\usepackage{amsthm}
\usepackage{mathtools}
\usepackage{physics}
\usepackage{xcolor}
\usepackage{graphicx}
\usepackage[left=23mm,right=13mm,top=35mm,columnsep=15pt]{geometry} 
\usepackage{adjustbox}
\usepackage{placeins}
\usepackage[T1]{fontenc}
\usepackage{lipsum}
\DeclareGraphicsExtensions{.pdf,.png,.jpg,.eps}
\usepackage{csquotes}
\usepackage{hyperref} 
\usepackage{xcolor}
\usepackage{soul}

\hypersetup{
	colorlinks=true,
	linkcolor=blue,
	citecolor=blue,
	filecolor=blue,
	urlcolor=blue,
}
\usepackage{color}
\usepackage{ulem}

\newcommand\old{\bgroup\markoverwith{\textcolor{ForestGreen}{\rule[0.5ex]{2pt}{0.8pt}}}\ULon}
\usepackage{subfiles} 

\begin{document}

\preprint{AIP/123-QED}

\title%
{All-optical bubble trap for ultracold atoms in microgravity}

\author{R. Veyron}
\affiliation{%
ICFO - Institut de Ciencies Fotoniques, The Barcelona Institute of Science and Technology, 08860 Castelldefels, Barcelona, Spain
}%

\author{C. Métayer}
\affiliation{ 
LP2N, Laboratoire Photonique, Numérique et Nanosciences, Université de Bordeaux-IOGS-CNRS:UMR 5298, rue F. Mitterrand, Talence F-33400, France
}

\author{J.B. Gerent}
\affiliation{Department of Physics and Astronomy, Bates College, Lewiston, ME 04240, USA}%

\author{R. Huang}
\author{E. Beraud}
\affiliation{ 
LP2N, Laboratoire Photonique, Numérique et Nanosciences, Université de Bordeaux-IOGS-CNRS:UMR 5298, rue F. Mitterrand, Talence F-33400, France
}

\author{B.M. Garraway}
\affiliation{
Dept.\ of Physics and Astronomy, University of Sussex, Falmer, Brighton, BN1~9QH, UK}

\author{S. Bernon}
\author{B. Battelier}%
 \email{baptiste.battelier@institutoptique.fr}
\affiliation{ 
LP2N, Laboratoire Photonique, Numérique et Nanosciences, Université de Bordeaux-IOGS-CNRS:UMR 5298, rue F. Mitterrand, Talence F-33400, France
}%

 \homepage{https://www.coldatomsbordeaux.org/}

\date{\today}

\begin{abstract}
    In this paper, we present an all-optical method to produce shell-shaped traps for ultracold atoms in microgravity. Our scheme exploits optical double dressing of the ground state to create a short range strongly repulsive central potential barrier. Combined with a long range attractive central potential, this barrier forms the shell trap.
    We demonstrate that a pure spherical bubble, reaching the quasi 2D regime for standard atom numbers, could be formed from two crossed beams with a parabolic profile. An analytical study shows that the relevant characteristics of the trap depend on the ratio of the ground and excited state polarisabilities 
    and the lifetime of the excited state. As a benchmark, we provide quantitative analysis of a realistic configuration for rubidium ensembles, leading to a 250 Hz transverse confinement for a 35 \textmu{m} radius bubble and a trap residual scattering rate of less than 10 s$^{-1}$.
\end{abstract}
\maketitle


\section{Introduction}

The study of ultracold atomic (UCA) systems, or ultracold quantum gases has often been guided by exploring different trap geometries, dimensionalities and topologies since their time-evolution depends strongly on the trapping potential. Indeed, trap features are expected to translate into exotic behaviors of the quantum gas. While most trap geometries are 'flat', reaching for curved geometries greatly extends the accessible landscape of physical phenomena, especially when the trap is closed onto itself. For example, ring trap\cite{Morizot2006} (1D) and torus\cite{PhysRevA.75.063406} (2D) geometries are being investigated as matter-wave guides that can realize inertial sensors and gyroscopes \cite{Amico2021} or be tools in the measurement of quantum phases \cite{Morizot2006}. 

Other interesting 2D candidates, under extensive study for the past few years, are shell-shaped traps \cite{garraway2016review,perrin_advamop_2017,Dubessy2025} on which we focus in this paper.
This specific trapping geometry is interesting for two main reasons. First, it allows the study of another part of the quantum physics underlying UCA systems. An extensive theoretical study of the quantum statistical properties and thermodynamic evolution of shell-shaped 2D Bosonic gases shows that they differ strongly from their 3D filled counterparts because of finite-size geometry dependent corrections \cite{Tononi2019,TONONI2024}. For example, Bose-Einstein condensation can be present before the Berezinskii–Kosterlitz–Thouless (BKT) transition (i.e.\ $T_{BKT} < T_{BEC}$), especially when $n_{2D} R^2$ is not too big (i.e.\  at small sphere radius $R$ or low atom density $n_{2D}$). Thus the range of temperatures for which a BEC can exist without superfluidity decreases when the size of the sphere or the atom number increases. In addition, UCA systems in shell traps present interesting properties such as dips in collective mode frequencies \cite{Sun2018,Padavic2017} or spontaneous vortex formation \cite{Turner2010,Vitelli2004}. They also constitute a useful platform to study Efimov physics in curved geometry \cite{Naidon_2017,DIncao_2018} and the physics of planetary atmospheres \cite{Saito2023}. 

Two distinct approaches to UCA
systems trapped in shells have been studied theoretically and demonstrated experimentally. The first approach relies on
adiabatic potentials created with radio-frequency (RF) dressed magnetic traps \cite{Zobay2001,garraway2016review,perrin_advamop_2017,Dubessy2025}. This idea has been implemented on Earth to trap quantum gases in 2D
shells \cite{Merloti2013}. But then the rotational symmetry is broken by the presence of the gravitational force. The realisation of closed shells with this method requires either
  small shells, or a compensation of gravity by optical \cite{Shibata2020,Zobay2001} or magnetic \cite{Guo2022} means, or a micro-gravity environment. 
As an example of the latter case, closed shells have been demonstrated onboard the International Space Station
(ISS) \cite{Aveline2020,Carollo2022} although other micro-gravity simulators could also be suitable. 
\citep{Condon2019,Lotz20171127,ZOEST2010,Becker2018} 
Challenges for RF dressed magnetic traps include
%
oblateness or prolateness of the shell potential, magnetic and RF noise and, 
in the case of a magnetic quadrupole geometry,  atom loss due to locations on the shell with vanishing RF coupling \cite{Guo2022}. Also, spatial 
inhomogeneities of the magnetic trap, inherent to atom chips, can lead to a shell radial vibration frequency which varies around the shell and in turn this can result
in a spatially varying zero-point energy\cite{Guo2022,Carollo2022} and hence additional forces around the shell.
A second approach uses immiscible dual-species BEC where the interspecies repulsive interaction ensures the formation of a closed shell of sodium atoms with its center filled with rubidium atoms \cite{Jia2022}. 
This method does not require a micro-gravity environment
, but is technically challenging and the shell radius is limited to less than a few tens of microns and with reduced tunability (due to the radius being controlled by the core atom number). 
In this approach 
there is also no freedom to apply another homogeneous magnetic field to take advantage of Feshbach resonances. 

In the present paper, we propose a third all-optical approach, to produce a fully spherical and closed bubble trap. Replacing the RF field of the dressed state technique by an optical one, an homogeneous Rabi coupling 
can be achieved by trapping the atoms in a dipole trap with a spatially uniform polarization. Furthermore, the spherical symmetry can be guaranteed using spatial light engineering such as painting potentials using fast-scanning acousto-optic deflectors (AOD), spatial light modulators (SLM) or digital micromirror device (DMD)\cite{Amico2021}. 
Additionally, our all-optical alternative method allows independent control of the full set of experimental parameters. More specifically it has both the advantages of inherently decoupling the shell trapping and the tuning of Feshbach resonances, and allows independent control of the radius of the sphere and its thickness. Among the technical advantages, this method preserves the optical access and allows for quasi instantaneous extinction of the trapping field. 

In the next Section~\ref{sec:bubble_optical_DDS} of this paper, we present the general idea of the optical approach to a bubble system state, which has similar aims and motivations as the RF dressed bubble state.
%
We then develop our model of a doubly-dressed quantum bubble state, by calculating the engineered forces to create the repulsive inner barrier while keeping a trapping force on the outer edge of the bubble-shaped trap. Then in Section~\ref{sec:Application to Rubidium} we develop a quantitative study applied to 
rubidium. Residual spontaneous emission is not negligible due to the natural linewidth of the intermediate state. We present a strategy using an additional laser, which brings another degree of freedom and allows a mitigation of the
photon scattering rate. In the subsection~\ref{sec:ExpeImplement} we propose an experimental implementation with realistic considerations. 
The paper then concludes with a discussion in Section~\ref{sec:Discussion}.

\section{Model: Double Dressed states to create an all optical bubble}
\label{sec:bubble_optical_DDS}

\subsection{ Double dressed states model}
\label{subsec:ToyModel}

\begin{figure}[!htb]
\centering
  \includegraphics[width=0.4\textwidth]{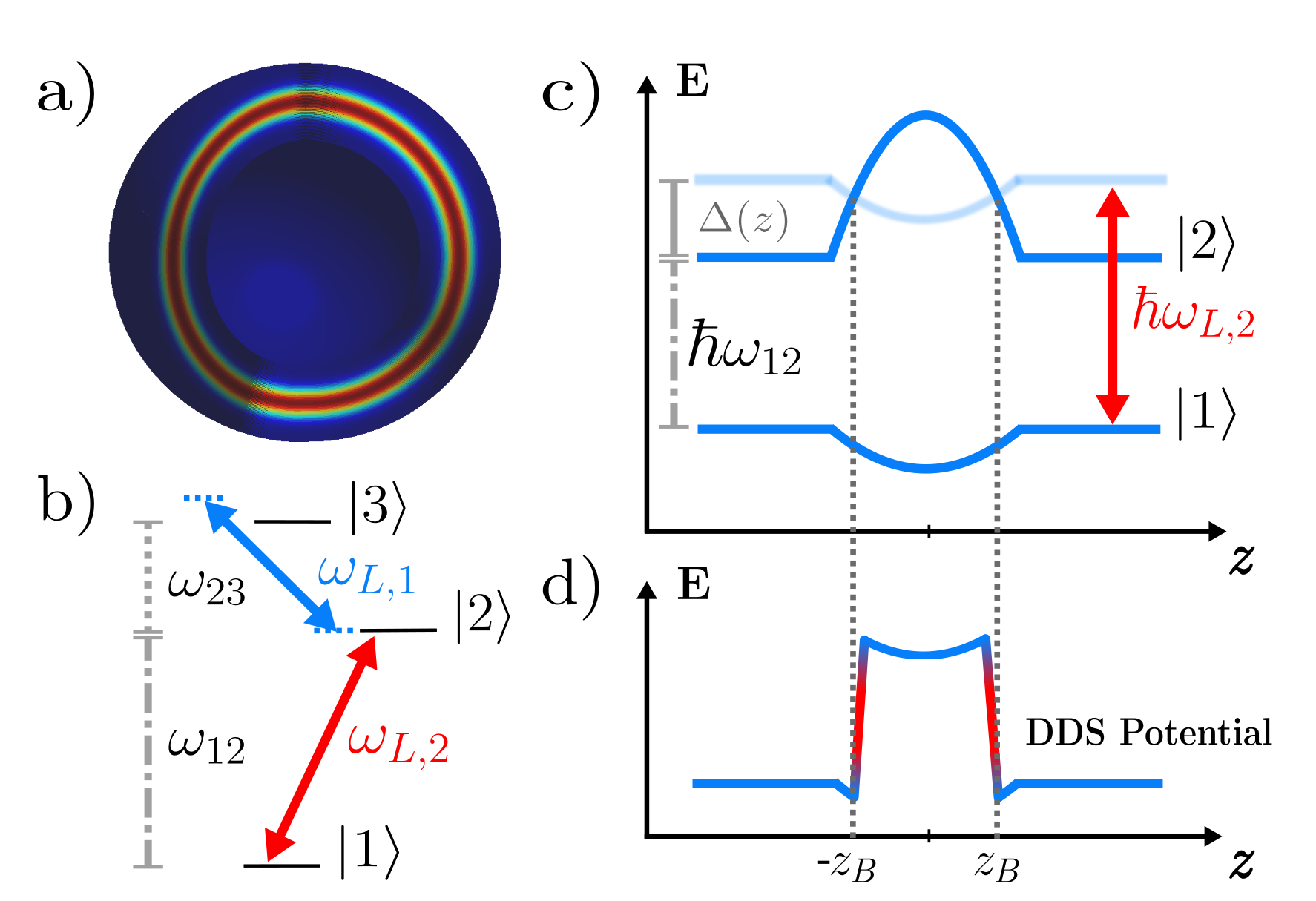}
\caption{ a) 3D representation of a bubble-shaped potential b) The three-level system is dressed by two lasers, the first one $\omega_{L,1}$ is blue-detuned on the $\ket{2} \rightarrow \ket{3}$ transition and red-detuned compared to $\ket{1} \rightarrow \ket{2}$. 
c) Eigen-energy profiles of $\ket{1}$ and $\ket{2}$ states along the $z$ direction when dressed by $\omega_{L,1}$ alone. The two shifts are parabolas of opposite sign. When dressed a second time by a laser radiation at $\omega_{L,2}$ this leads to d), \textit{i.e.}, a double well at the crossing of the double dressed state, corresponding to a sphere in 3D.  
The final doubly-dressed state potential comes from the relative probability for the atom to be in the state $\ket{1}$ (blue) or $\ket{2}$ (red) spatially modulated by the effective detuning $\Delta(r)$. 
}

\label{fig:DoubleDressedStates} 
\end{figure}

Our method to produce quantum bubbles is inspired by double dressed states (DDS) which have been studied in our group to trap rubidium atoms close to a surface \cite{Bellouvet2018} and for super-resolution imaging in optical lattices \cite{VeyronPRXq2024}. The principle is similar to the RF dressing approach\cite{Zobay2001,garraway2016review,perrin_advamop_2017,Dubessy2025}, but in the optical range, 
and is valid for any three-level system with decreasing transition frequencies ($\omega_{32}<\omega_{21}$) as illustrated in Fig. \ref{fig:DoubleDressedStates}(b). We depict the principle in 1D (Fig. \ref{fig:DoubleDressedStates}(c,d)) corresponding to the radial direction of the sphere. 
We consider a first laser radiation blue-detuned and far from resonance compared to $\ket{2} \rightarrow \ket{3}$ transition and red-detuned compared to $\ket{1} \rightarrow \ket{2}$ transition $\omega_{32}\ll \omega_{L,1}\ll \omega_{21}$. We assume that this radiation has a parabolic transverse profile with maximal intensity at $r=0$. The intermediate state  $\ket{2}$ sees a repulsive potential from $r=0$ while the atoms in the ground state  $\ket{1}$ are trapped at $r=0$. As shown in Fig. \ref{fig:DoubleDressedStates}(c), the $\ket{1} \rightarrow \ket{2}$ transition frequency  $\omega_{0}(r)$ therefore depends on the radial position $r$. 
Double dressing is then achieved by applying a second laser radiation at frequency $\omega_{L,2}$ tuned close to the free space  $\ket{1} \rightarrow \ket{2}$ transition $\omega_{21}$. The resulting AC Stark shift on the state $\ket{1}$ is governed by the spatially-dependent detuning $\Delta(r) = \omega_{L,2}-\omega_{0}(r)$ (see Fig. \ref{fig:DoubleDressedStates}(c)), which cancels at position $r_{\rm bubble}$. This position directly depends on the laser frequency $\omega_{L,2}$. For positions  $|r| < |r_{\rm bubble}|$, the detuning  $\Delta(r)$ is positive (blue) and the doubly-dressed potential expels the atoms away from the center of the laser beam $\omega_{L,1}$. As depicted in Fig \ref{fig:DoubleDressedStates}(d) a trap is then formed by the doubly-dressed state potential. 

To extend the solution to 3D, we propose to produce a maximum of intensity in three dimensions using a set of non-interfering crossed beams (incoherent or time-averaged) with parabolic profiles. As shown below, in the absence of gravity, this produces a perfectly spherical shell as shown in Fig. \ref{fig:DoubleDressedStates}(a).

\subsection{Analytical derivation}
\label{subsec:DDSModel}

In this section, we derive from the doubly-dressed state formalism 
the optical potential of the shell trap and give an analytical expression of the geometrical parameters of a spherically symmetric trap in the case of parabolic 3D energy shifts.
For simplicity, we consider a three-level atom as in Fig. \ref{fig:DoubleDressedStates}(b) interacting with two distinct laser fields. The laser $1$ has an angular frequency  $\omega_{L,1}$ which is far off-resonance 
from the transition $\ket{2}\leftrightarrow \ket{3}$. The laser $2$ has an angular frequency $\omega_{L,2}$ which is off-resonant but close to the transition $\ket{1}\leftrightarrow \ket{2}$ in order to shift strongly the state $\ket{2}$ while keeping the state $\ket{1}$ minimally shifted. The total dipole force is given by \cite{Dalibard1985}:
\begin{eqnarray}
\begin{aligned}
    \vec{f}_\text{dip}(x,y,z) &= -\vec{\nabla} U (x,y,z) \nonumber\\
    &= - \Pi_1(x,y,z) \vec{\nabla} U_1(x,y,z) - \Pi_2(x,y,z) \vec{\nabla} U_2(x,y,z)
    ,
    \label{sec:two_level}
\end{aligned}
\end{eqnarray}
where $\Pi_i$ and $U_i$ are the population and potential energy of the state $\ket{i}$ dressed by the laser field $\omega_{L,1}$. Here, we neglect the population in the third level leading to population conservation such that $\Pi_1 + \Pi_2 = 1$. The population $\Pi_2$ transferred to the dressed state $\ket{2}$ by the laser 2  is derived from the Optical Bloch formalism as  $\Pi_2(\Delta(x,y,z))=\frac{s}{1+s+4(\Delta(x,y,z)/\Gamma)^2}$ where $\Gamma$ is the spontaneous emission rate of the excited state $\ket{2}$, $s$ the saturation parameter induced by laser 2 on the $\ket{1}\leftrightarrow \ket{2}$ transition,  $\Delta(x,y,z)=\Delta_\text{0}+(U_1(x,y,z)-U_2(x,y,z))/\hbar$ is the spatially-dependent detuning, and $\Delta_\text{0}=\omega_{L,2}-\omega_\text{21}$ is the laser detuning with respect to the bare $\ket{1}\leftrightarrow \ket{2}$ transition. As shown in \ref{Meth:DDSderivation},
the total doubly-dressed-state potential at position $\mathbf{r}$ is then analytically obtained by a spatial 1D integration of the dipole force :
\begin{equation}
    U(\mathbf{r}) = U_\text{1}(\mathbf{r}) - \frac{\hbar \Gamma s}{4\sqrt{1+s}} \left(  \atan\left[\frac{2\Delta(\mathbf{r})}{\Gamma \sqrt{1+s}}\right]- \atan\left[\frac{2\Delta_{0}}{\Gamma \sqrt{1+s}}\right]  \right)
    .
    \label{eq:general_integral_DDS_1d_4}
\end{equation} 

To create a 3D spherical shell trap, we consider the case of a 3D spherically symmetric parabolic potential which is experimentally feasible (see section \ref{sec:ExpeImplement}):
\begin{equation}
    U_\text{i}(x,y,z) = U_\text{i}(r) = \tilde{U}_{\text{i}} \left(  1-\frac{r^2}{R_{i}^2} \right) 
    \label{eq:potential_DDS3D_SpheCoor_HO_g}
\end{equation}
where $i\in \{1,2\}$ is the dressed state index and $R_{i}$ is the parabolic radius of state $i$. To reach a fully symmetric bubble, we consider the case of identical potentials $R_1=R_2=R_0$. 
Moreover, to shape a bubble, the laser 1 needs to be blue-detuned for the transition $\ket{2}\leftrightarrow \ket{3}$ and red-detuned for $\ket{1}\leftrightarrow \ket{2}$ leading to the two potentials  $\Tilde{U}_1 <0$ and $\Tilde{U}_2 > 0$ having opposite sign.

\begin{figure}[!htb]
    \centering
    \includegraphics[width=1\linewidth]{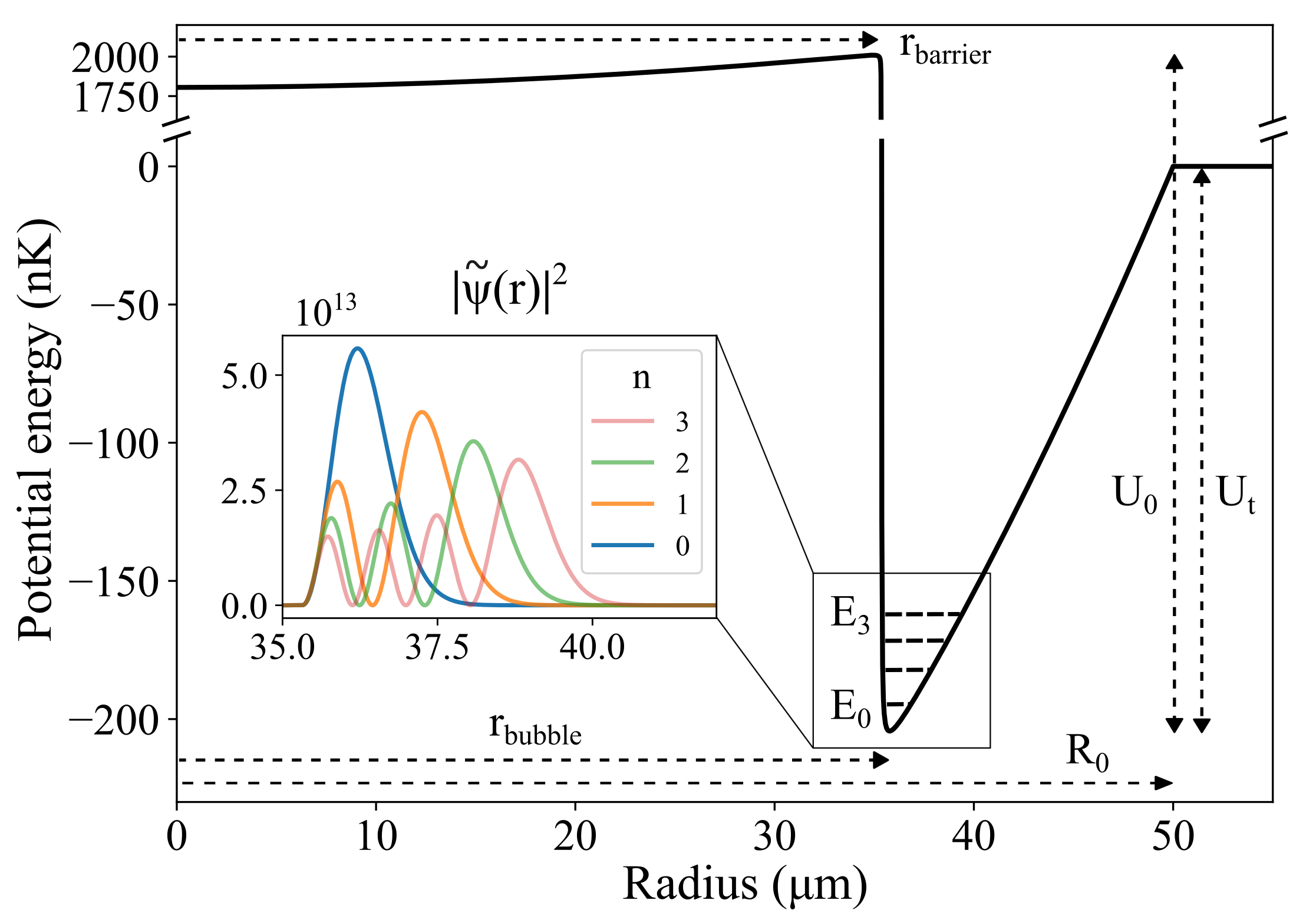}
    \caption{Doubly-dressed state potential in the radial direction leading to a bubble-shaped trap with spherical symmetry in 3D. The inset shows the probability densities $|\tilde{\Psi}(r)|^2$ 
      of the first four eigenstates
    without interaction and normalized in 3D in spherical coordinates.} 
    \label{fig: generic bubble potential}
\end{figure}

In such a configuration, we can analytically express relevant parameters of the shell traps which are schematically represented in Fig. \ref{fig: generic bubble potential}. While the exact solutions are given in \ref{Meth:DDSderivation}, we focus here on the low saturation limit approximation ($s\ll1$) which is relevant for experimental implementation. In this limit, the shell trap radius $r_{\rm bubble}$ as defined by the local minimum of potential energy and the central barrier radius $r_{\rm barrier}$ are given by : 
\begin{equation}
    r_\text{bubble}^2   = (1-\eta) R_0^2  + \frac{\hbar\Gamma}{2\Tilde{U}_2} R_0^2 \sqrt{\beta -1}
    \label{eq:Rbubble}
\end{equation}
\begin{equation}
     r_\text{barrier}^2  = (1-\eta) R_0^2  - \frac{\hbar\Gamma}{2\Tilde{U}_2} R_0^2 \sqrt{\beta -1}
    \label{eq:rbar} 
\end{equation}
where $\eta=\hbar \Delta_{0}/(\Tilde{U}_2-\Tilde{U}_1) $ is the laser detuning in units of the differential light shift and $\beta=s \Tilde{U}_2/2|\Tilde{U}_1|\gg 1$ is a dimensionless parameter. The condition on the detuning to create a bubble is $\eta\in [ 0,1]$.

Given these positions, we can define the external trap depth as 
$U_t = U(\pm \infty) - U(r_\text{bubble}) = - U(r_\text{bubble}) $ and the central barrier height $U_0 = U(r_\text{barrier}) - U(r_\text{bubble})$ as the energies required to escape towards respectively the outer and inner bounds:
\begin{eqnarray}
\begin{aligned}
   U_t &\approx  \eta |\Tilde{U}_1| -  \frac{s \hbar \Gamma}{2\sqrt{\beta}} \\
   U_0 &\approx s \hbar \Gamma \frac{\pi}{4} \left( 1- \frac{4}{\pi\sqrt{\beta}} \right) 
   . \label{eq:U0}
\end{aligned}
\end{eqnarray}
We observe from these expressions that the excited state contribution $s\hbar\Gamma$ involving the linewidth and the saturation parameter 
should be kept quite large 
to create a significant central barrier height. The consequences are a reduction of the trap depth and a residual scattering rate. Such a limit is characterized by the residual scattering time at the bubble position: 
\begin{equation}
    \tau_d = \frac{1}{\Gamma \cdot \Pi_2(r_{\text{bubble}})} = \frac{ 2 + \Tilde{U}_2/\abs{\Tilde{U}_1} }{\Gamma} \label{eq:lifetime}
    .
\end{equation}
This expression clearly emphasizes the role played by the relative energy shift $\Tilde{U}_2/\abs{\Tilde{U}_1}$ which should be engineered to be as large as possible.

In the radial direction, the bubble is strongly anharmonic due to the sharp potential barrier. The energy spectrum cannot be described by a single "trap frequency" as in the harmonic case. To define a useful parameter to study the dimensionality of the quantum gas dynamics, we use the energy difference between the first two eigenstates in the case without interactions. We approximate the bubble trap by a half-wedge potential, where the potential varies linearly above the bubble radius. We find that the energy spectrum of the full DDS potential agrees well with this approximated well-wedge potential (see \ref{App:states_without_interaction} for the full comparison and details). 
Then the energy difference between the first two states in the well-wedge potential is: 
\begin{equation}
    \frac{\Delta E_{1,0}}{\hbar}\approx 2.22 \left( \frac{ \tilde{U}_1^2 r_\text{barrier}^2 }{ m \hbar R_0^4 } \right)^{1/3} 
    .
    \label{eq:trap_frequency_from_spectrum}
\end{equation}

Expressions (\ref{eq:U0}-\ref{eq:trap_frequency_from_spectrum}) globally define the optimal choice of atomic species and laser configuration :
\begin{itemize}
 \item The atomic species should have a narrow linewidth $\Gamma$ but not too narrow to maintain a reasonable central barrier height in the low saturation limit (Eq. \eqref{eq:U0}).
 \item Long lifetime will be reached for atomic species with a strong dipole element on the $\ket{2}\leftrightarrow \ket{3}$ transition that will increase $\Tilde{U}_2/|\Tilde{U}_1|$ (Eq. \eqref{eq:lifetime}). 
 \end{itemize}


Finally, our approach allows the production of a versatile bubble with an accurate control of the radius and the radial confinement. The bubble size is controlled by the radius $R_0$ of the parabolic profile 
(Eq.~(\ref{eq:potential_DDS3D_SpheCoor_HO_g}))
and the relative detuning $\eta$ of the second dressing laser compared to the light shift $\tilde{U}_2$ (see Eq. \eqref{eq:Rbubble}). The tightness of the trap, which is related to the thickness of the bubble, is mainly determined by the energy difference \eqref{eq:trap_frequency_from_spectrum} which scales as the intensity of the first dressing laser and the inverse of the radius of the parabola $\Delta E_{0,1}\propto (\tilde{U}_1/R_0)^{2/3}$ according to Eqs. \eqref{eq:Rbubble} and \eqref{eq:trap_frequency_from_spectrum}.


\section{Application to rubidium $^{87}Rb$}
\label{sec:Application to Rubidium}
To study a practical implementation, we choose rubidium $^{87}Rb$ which can be efficiently cooled and possess excited transition in the telecom frequency band where high power lasers are accessible. 
Moreover, from a technology point of view, ultracold rubidium experiments were already demonstrated in the International Space Station \cite{Aveline2020} and in microgravity platforms \cite{ZOEST2010,Becker2018,Condon2019}.

\subsection{Equivalent three-level system}
\label{sec:Rb3LS}
The control scheme presented in section \ref{sec:bubble_optical_DDS} relies on the existence of a three-level system. Here we propose applying the method on the following states: $\ket{1}=\ket{5S_{1/2}, F=2, m_F=2}$, $\ket{2}=\ket{5P_{3/2}, F=3, m_F=3}$, $\ket{3}=\ket{4D_{5/2}}$ (or $\ket{4D_{3/2}}$).
Since the state $\ket{3}$ is being probed off resonance the details of its hyperfine and Zeeman manifold are not relevant. When probed by a laser radiation close to 780 nm with a $\sigma_+$ polarisation, $\ket{1}\leftrightarrow\ket{2}$ forms a closed two-level system. Due to selection rules of the electric dipole interaction, this transition stays closed when dressed by a 
$\pi$-polarized 1529-nm radiation. 
The experimental realization of such a configuration of the radiation polarization is shown in Fig. \ref{fig:Setup final config 3D}.

\subsection{Optical parameters to produce a bubble-shaped trap}
\label{sec:BubbleParameters}

We give here a set of optical parameters to produce an all-optical shell-shaped trap. 
The radius of the parabolic laser beam is fixed to $R_0=50$ \textmu{m}, which allows the production of the largest bubble 
related to the maximum optical power $P_1=20 $ W currently available with telecom technology.
This power leads to a strong lightshift of the intermediate state $\tilde{U}_2/\hbar \approx 5918\Gamma \approx 36$ GHz at 0.26 nm detuning. 
This strong light shift limits the spontaneous emission suffered by the double dressed state at the location of the wave-function (Eq. \eqref{eq:lifetime}). 

The second dressing laser requires an homogeneous Rabi coupling over the full atomic sample. A single circularly-polarised laser at $\lambda_2=780$ nm is sent on the atomic sample with a spatially homogeneous intensity and a saturation parameter $s = 0.01$.  The relative detuning $\eta = 0.5$ is fixed at half the maximum of the light shift $\tilde{U}_2$ 
leading to a radius $r_\text{bubble} = 35.8$ \textmu{m}. 

Incidentally, the first dressing laser also creates a light shift of the ground state $\tilde{U}_1/{\hbar} = -1.64 \Gamma$ 
corresponding to the usual dipole trap and playing here the role of the attractive potential on the external edge of the bubble trap. 

\subsection{Lifetime of the optical bubble}
\label{sec:Lifetime optimisation}

Without specific care, the short lifetime of the intermediate state of rubidium $^{87}Rb$, $\Gamma^{-1}=27$ ns, leads to significant spontaneous emission. To reach experimental timescales of 0.1 to 1~s, corresponding to typical low temperature dynamics, the spontaneous rate should be mitigated by seven orders of magnitude. 
According to Eq. \eqref{eq:lifetime}, the ratio of the light shift of the ground and intermediate state $\tilde{U}_2/\abs{\tilde{U}_1}$ can counter-balance the spontaneous emission.

\subsubsection{Simplest strategy : single laser}

\begin{figure}[!htb]
    \centering
    \includegraphics[width=1\linewidth]{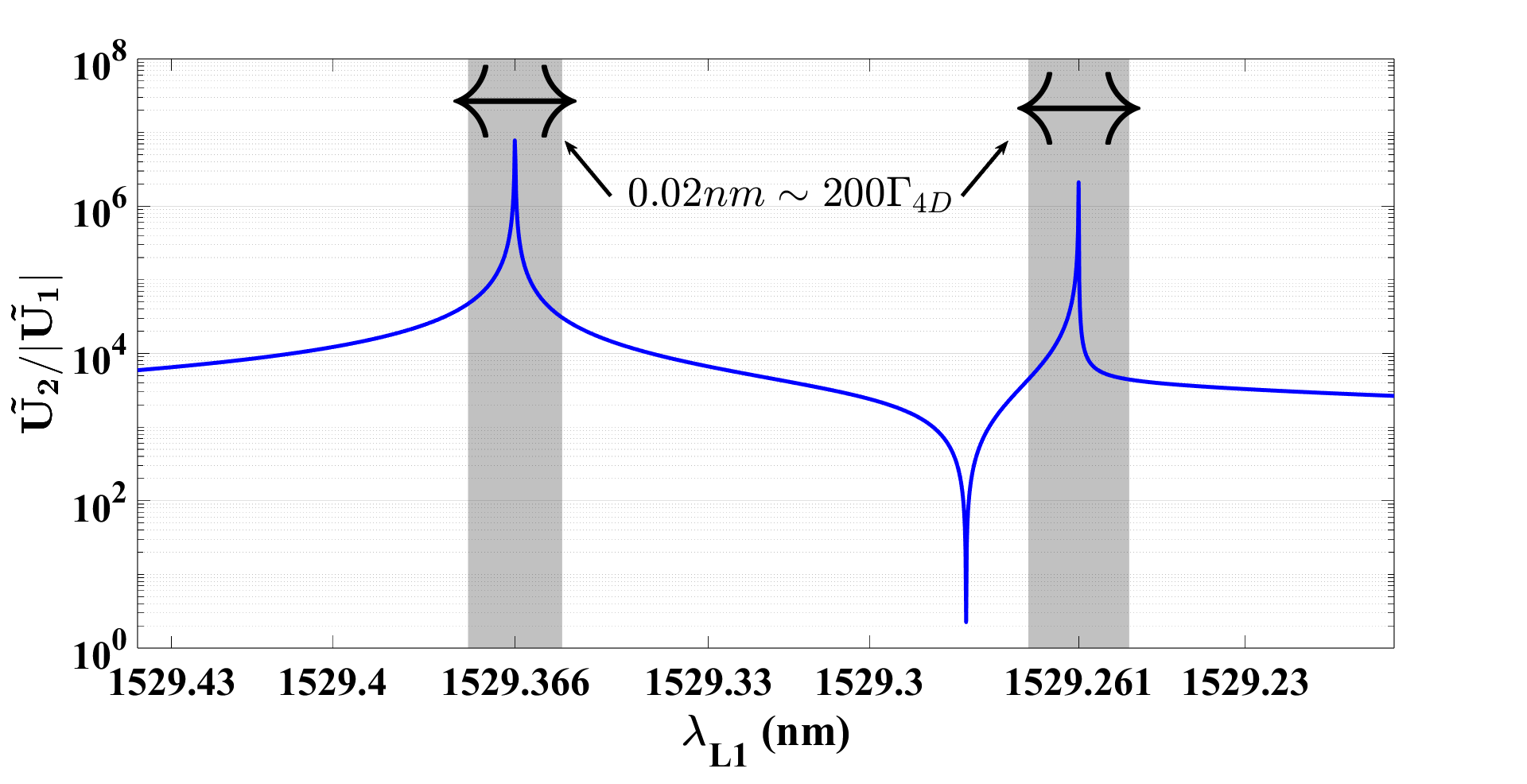}
    \caption{Ratio of the lightshifts $\tilde{U}_2/\abs{\tilde{U}_1}$ at low intensity $I=3800$ W.cm$^{-2}$  versus the wavelength of the first dressing laser $\lambda_{L,1}=2\pi c/\omega_{L,1}$. The two gray shaded areas show the zone where the laser is close to the two resonances $\ket{5P_{3/2}, F=3, m_F=3}\rightarrow \ket{4D_{5/2}}$ and $\ket{5P_{3/2}, F=3, m_F=3}\rightarrow \ket{4D_{3/2}}$. 
    }
    \label{fig: ratio polarizabilities}
\end{figure}

If light from a single laser is used to control both light shifts, in the low power regime, the ratio $\tilde{U}_2/\abs{\tilde{U}_1}$ corresponds to the ratio of polarisabilities for state $\ket{1}$ and $\ket{2}$. 
Figure \ref{fig: ratio polarizabilities} shows the value of this ratio for a $\pi$-polarized radiation tuned in the wavelength range $1529.31 \pm 0.12$ nm. Details of the corresponding calculation are given in \eqref{App:LightShifts}. As shown on the figure, the large ratio $\tilde{U}_2/\abs{\tilde{U}_1}>5\times 10^4$ (gray shaded area) can only be reached very close to resonance where it is still quite limited and out of the domain of validity for the DDS formalism. 

Typically, with a laser beam at 1529.34 nm between the two transitions we get $\tilde{U}_2/\abs{\tilde{U}_1} \approx 10^4$ leading to a diffusion time $\tau_d \approx 0.3$ ms. We notice that this calculation in the low intensity regime represents a best case scenario because it overestimates the lightshift ratio one would obtain by exact diagonalisation in the high intensity regime.

\subsubsection{Increasing the lifetime with a compensation laser}
To add another degree of freedom, we consider using a third laser at wavelength $\lambda_3 = 770$ nm blue-detuned for the $D_2$ transition and with  a radius identical to the harmonic laser beam profile $R_0^{(770)}=R_0^{(1529)}$. This wavelength is chosen to be detuned enough to avoid additional spontaneous emission but close to the $D_2$ transition, so as to affect mostly the states $\ket{1}$ and $\ket{2}$. Additionally, on a technological point of view, this wavelength corresponds to robust standard laser technology producing the required power level. This additional laser allows us to increase the ratio $\tilde{U}_2/\abs{\tilde{U}_1}$ and thus reduce the photon scattering rate. 

We choose this strategy for the following parts of the article and the experimental realization of such configuration is detailed in section \ref{sec:ExpeImplement}.

\subsection{Energy scales in the all-optical quantum bubble}
\label{sec:EnergyScale}

\begin{figure}[!htb]
    \centering
    \includegraphics[width=1.1\linewidth]{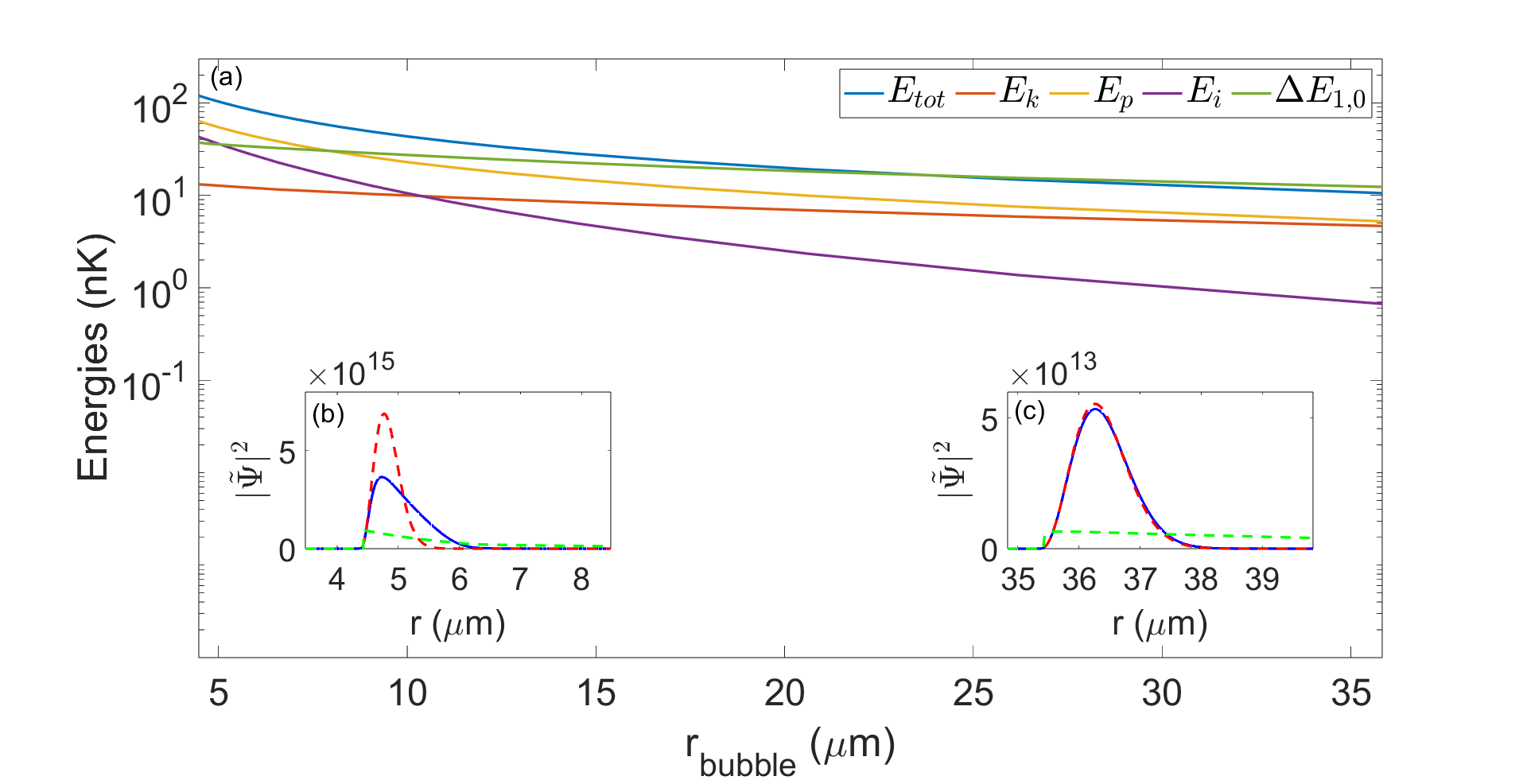}
    \caption{Kinetic, potential, interaction and total energy versus the bubble radius which is controlled by varying the radius of the parabolic beam at constant laser intensity. Wave-function probability density
    calculated using the numerical model for a radius $r_{\rm bubble}=4.5$ \textmu{m} (b) and $r_{\rm bubble}=36$ \textmu{m} (c) 
    respectively, illustrating a cross-over from the Thomas-Fermi regime (dashed green line) to the non-interacting regime (dashed red line). }
    \label{fig:EnergiesversusRadius}
\end{figure}

Fig. \ref{fig:EnergiesversusRadius} presents the different energies involved in the system calculated numerically for $N_\text{at} = 10^5$. 
We will now compare the kinetic, potential and interaction energies with the energy splitting between the two first levels calculated in the non-interacting regime (see Eq. \eqref{eq:trap_frequency_from_spectrum}). The bubble radius is controlled by varying the radius of the parabolic potential $R_0$ at constant intensity, the optical power being increased while $R_0$ decreases. For short radius  ($r_{\rm bubble}< 10$ \textmu{m}), the interaction energy is higher than the kinetic energy leading to the Thomas-Fermi regime. For a larger bubble radius, the interaction energy decreases to become negligible, leading to the non-interacting regime. This is confirmed by the spatial profile of the wave-function calculated numerically by solving the Gross-Pitaevskii equation (see \ref{App:Numerical model}) compared to the expected wave-function profiles in the Thomas-Fermi and non-interacting regime respectively (insets of Fig. \ref{fig:EnergiesversusRadius}). For $N=10^5$ and a radius $r_{\rm bubble} > 20$ \textmu{m} we notice the interaction energy starts to become negligible compared to the energy difference between the ground state and the first excited state. For low temperature satisfying $T< \Delta E_{1,0} \approx k_B\cdot12$ nK $\approx h\cdot 250$ Hz, this non-interacting regime matches with the quasi-2D regime, confirmed by the calculation of the chemical potential  in \ref{App:Numerical model}: 
\begin{equation}
        \mu = \left ( \frac{3 N a \hbar \Delta E_{1,0}}{2\sqrt{2} \sqrt{m} r_{\rm bubble}^2} \right ) ^{2/3}\label{eq:chemicalPot}
        .
\end{equation}

\begin{figure}[!htb]
    \centering
    \includegraphics[width=1\linewidth]{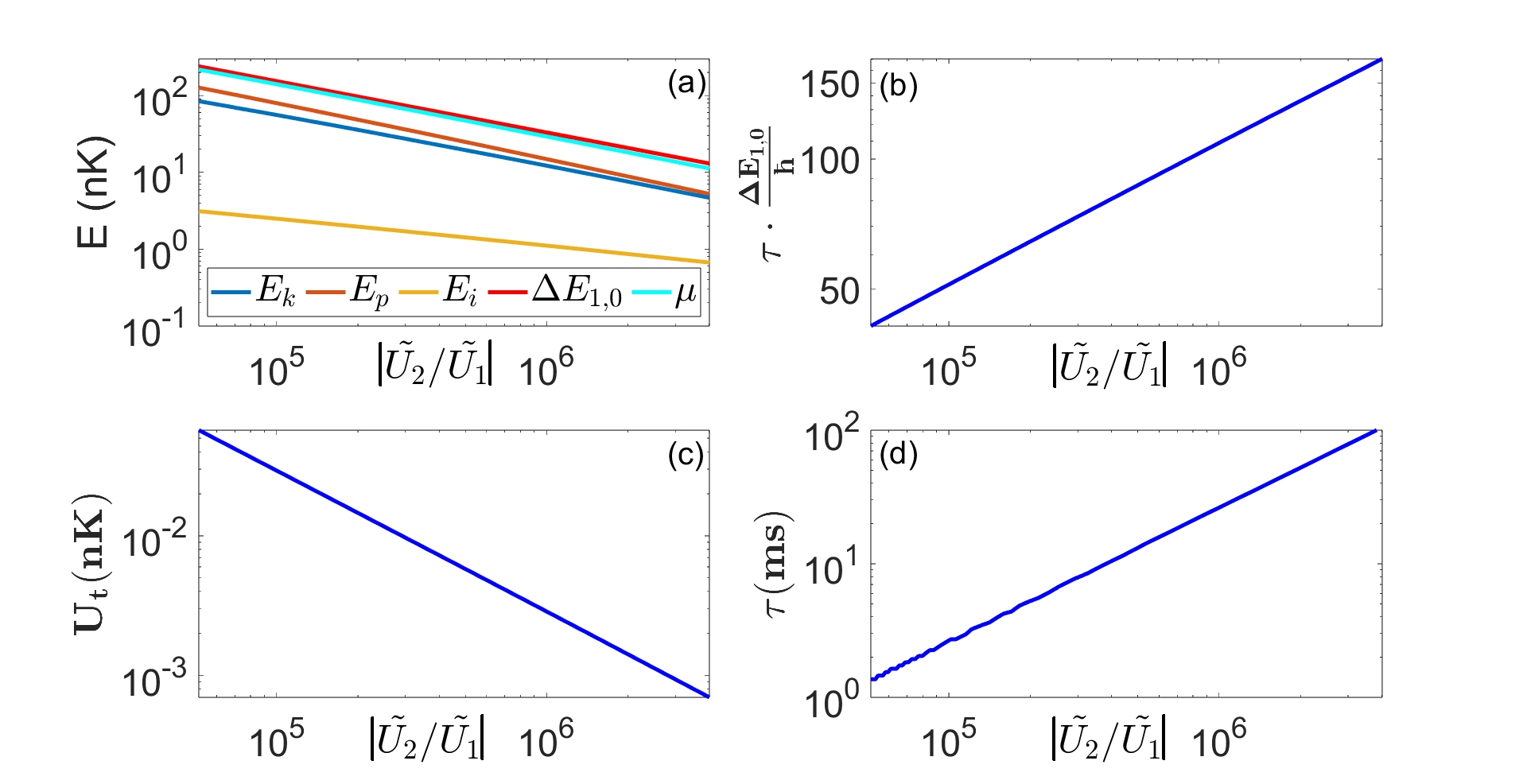}
    \caption{Parameters of the quantum bubble trap versus the lightshift ratios:  (a) Kinetic $E_{k}$, potential $E_{p}$ and interaction energy $E_{i}$; (b) product of the lifetime $\tau$  times the trap frequency $\omega$ as the figure of merit; (c) Final trap depth $\tilde{U}_t$; (d) diffusion lifetime due to spontaneous emission. 
    }
    \label{fig : variation laser power 2 1D 2 laser gaussian 1529 and 770}
\end{figure}

Using the relevant parameters of section \ref{sec:BubbleParameters} with a bubble radius $r_{\rm bubble}= 35$ \textmu{m}, a numerical simulation is presented on Fig. \ref{fig : variation laser power 2 1D 2 laser gaussian 1529 and 770} while varying the laser intensity of the laser at  $\lambda_{L,3}=770$ nm.  We confirm that the interaction energy $E_\text{i}$ is much lower than the other energies of the system for $N_\text{at}=10^5$ atoms (see Figure \ref{fig : variation laser power 2 1D 2 laser gaussian 1529 and 770} (a)). Moreover, we checked with our numerical model that the atom number can be increased until $N_\text{at}=2\cdot10^5$ while maintaining the system in the quasi-2D regime (see \ref{App:Numerical model}). 

Figure \ref{fig : variation laser power 2 1D 2 laser gaussian 1529 and 770} shows the most important parameters of the bubble system for $N_\text{at}=10^5$ as a function of the light shift ratio $\tilde{U}_2/\abs{\tilde{U}_1}$. This ratio is controlled by the intensity of the 770 nm dipole trap laser. The lifetime of the bubble due to the diffusion rate (see Eq. \eqref{eq:lifetime}) increases linearly with this ratio until reaching a critical laser power $P_3 = 913$ mW for which the light shift of the ground state is fully cancelled, removing the attractive force and consequently destroying the bubble-shaped trap. At this critical point the diffusion time is equal to $\tau_d \approx 105$ ms corresponding to the average time for one atom to scatter one photon and is considered to be the major limitation of trap lifetime. Another consequence of the cancellation of the total light shift of the ground state is the strong reduction of the trap depth. The spatial profile of the double dressed state potential is calculated numerically from Eqs. \eqref{eq:general_integral_DDS_1d_4} and \eqref{eq:potential_DDS3D_SpheCoor_HO_g}. Taking into account the effect of the third laser $-U_t^{(3)}$, we check the trap depth $\tilde{U}_t=U_t-U_t^{(3)} \approx 200$ nK is higher than the critical temperature for Bose Einstein condensation $T_c=25 $ nK, given $N_\text{at}= 10^5$ in the non-interacting regime. The typical energy splitting ($\Delta E_{1,0} \approx 12$ nK) allows 27 bound states (see \ref{App:states_without_interaction}). 

Finally, we define a figure of merit as the product of the energy splitting of the lowest two states and the diffusion rate (Fig. \ref{fig : variation laser power 2 1D 2 laser gaussian 1529 and 770} (b)): this measures the trade-off between the lifetime and the experimental ability 
to be in the quasi-2D regime. It shows that the most favorable conditions are present while reaching the critical point at the cancellation of the ground state light shift.

\subsection{Experimental implementation of the spherical bubble}
\label{sec:ExpeImplement}
\begin{figure}[!htb]
    \centering
    \includegraphics[width=1\linewidth]{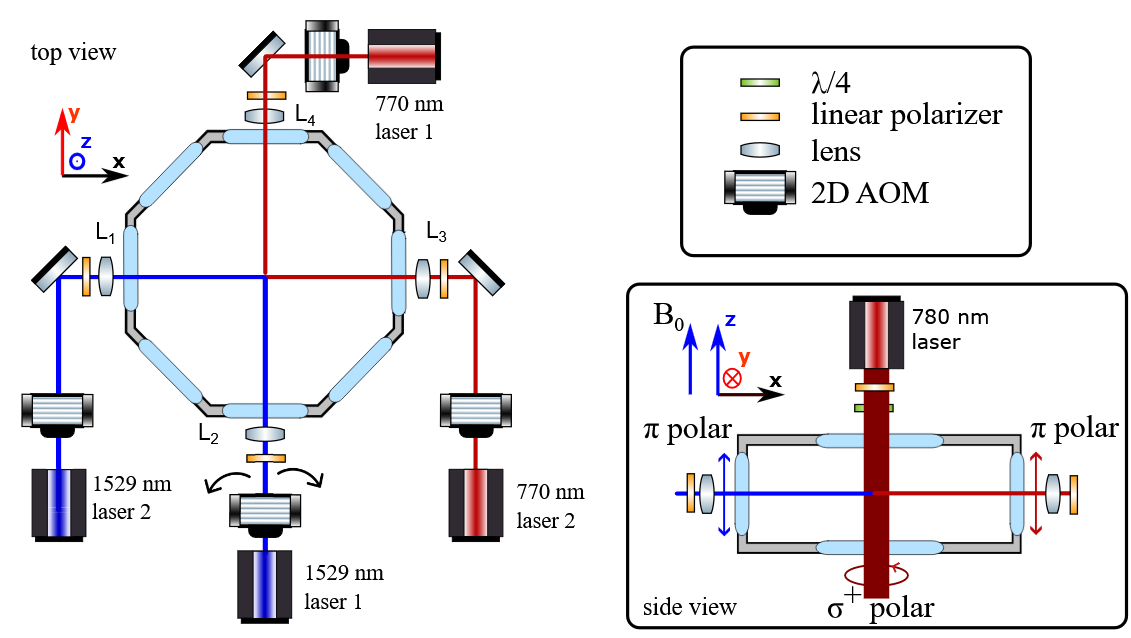}
    \caption{Schematic of an experimental implementation to produce the all optical shell shaped trap. The first dressing light is supplied by two independent 1529 nm laser sources. On each beam path, a pair of crossed acousto-optical modulators (2D AOM) creates the parabolic painted potential in 2 dimensions. This first painted potential is overlapped with an additional painted potential created by a second pair of laser sources emitting at $\lambda_{3}=770$ nm and associated to 2 other 2D AOMs. All the beams crossed at the center of an ultrahigh vacuum chamber. Some optical components are omitted for clarity. Inset: Side view showing the second dressing laser at $\lambda_{2}=780$ nm and the polarisation configuration of the different laser beams.}
    \label{fig:Setup final config 3D}
\end{figure}

\bigskip

We propose a possible experimental implementation of the all-optical shell-shaped trap to illustrate the feasibility of our proposal.  Figure \ref{fig:Setup final config 3D} shows the general architecture, including three laser sources at the wavelengths mentioned in section \ref{sec:Lifetime optimisation}. The parabolic profile of the first dressing laser at $\lambda_1 = 1529$ nm is produced using acousto-optical modulators to spatially modulate the beam leading to a painted potential \cite{Amico2021}. Two of these painted parabolic beams, propagating in the (x,y) plane, cross each other   perpendicularly at the center of the science chamber. Each 2D painted profile has an elliptical symmetry to provide the 3D spherical parabolic potential required to form the full bubble.
Practically, the high required power leads to a non-linearity of the lightshift with the laser intensity. This means that a parabolic beam profile does not produce a parabolic painted potential. Fortunately this can be easily compensated by adapting the modulation profile of the painted potential. 

The second dressing laser at $\lambda_2 = 780$ nm propagates along z. The spatial intensity of this beam needs to be homogeneous over the full atomic sample to guarantee an homogeneous dressing coupling. Additionally a special care is required concerning the polarisation of the different beams to reproduce the three-level system (see section \ref{sec:Rb3LS}). More specifically, the polarisation of the two parabolic beams need to be linear, both along z. It is thus necessary to have two different lasers with no coherence between them to avoid any interference between the two crossed painted potentials. A static magnetic field $B_0$ sets the quantification axis along the z-axis, parallel to the propagating axis of the second dressing laser. 
The laser emitted at $\lambda_3 = 770$ nm is used to increase the light shift ratio $\tilde{U}_2/|\tilde{U}_1|$. This laser is overlapped with the first dressing laser in a contra-propagating configuration and yields the same painted potential spatial profile.

Our method relies on an ideal compensation for the ground state light shift $\tilde{U}_1$ over the full sphere. Indeed, a misalignment or a difference in the radius of the two painted potentials at $\lambda_1$ and $\lambda_3$ will be detrimental for two reasons. The first effect is a reduction of the lifetime. The second effect is the destruction of the bubble symmetry. 
It is critical because  the final light shift of the ground state is engineered by a rejection of two orders of magnitude of the initial light shift, due to the first dressing laser $U_2(r)=U^{(1)}_2(r)-U^{(3)}_2(r)\approx U^{(1)}_2(r)/100$. For a radius $R_0=50$ \textmu{m}, the required relative precision of the alignment of the two lasers $1$ and $3$ is of the order of 50 nm. Overlapping the two painted potentials with such a high precision seems very challenging.  Yet we emphasize that painted potentials allow complex optimization processes. Discussing the experimental alignment protocols of the set-up is out of the scope of this paper. 

\section{Conclusions}
\label{sec:Discussion}
We have presented a strategy using exclusively atom-light interaction to produce a bubble-shaped trap for ultra-cold atoms. Our method relies on creating a doubly-dressed state potential associated to spatial engineering of ac Stark shifts in a three-level atom, while compensating the gravity potential in microgravity or using an external force \cite{Shibata2020}. 
Our technique favorably allows the creation of a fully spherical trap and can produce a shell-shaped ultracold gas in the quasi-2D regime for a large range of parameters. By switching the detuning of the second dressing laser and the spatial modulation amplitude of the painted potential, the bubble radius and radial stiffness can be tuned easily until reaching the thin-shell limit, where the width of the radial wave-function is small compared to the bubble radius. This regime has already been the subject of numerous theoretical studies \cite{Padavic2017,TONONI2024}.

We have applied our method to rubidium atoms for which the technology has already been validated in microgravity environments. The main limitation comes from the atomic natural linewidth leading to a significant reduction of the lifetime due to photon scattering. 
This effect is mitigated using an extra laser 
allowing an increase in the lifetime of the ultracold atoms in the bubble shaped trap to above 100 ms. 

Our scheme relies on the ability to produce harmonic potentials, but the method can be adapted to different geometries. Practically, these harmonic potentials can be realized with Gaussian beams but that would
restrict the bubble radius to small values to maintain the harmonic approximation, which is mandatory to create a spherically-symmetric bubble.

A loading procedure of the optical bubble-shaped trap similar to RF dressed traps can be envisioned. At the beginning of the sequence, all three lasers would be turned on at the nominal power, but with a detuning of the second dressing laser settled "off resonance" $\eta > 1$, leading to a classical 3D potential trapping a Bose Einstein Condensate. While decreasing $\eta<1$, the inner barrier would start to appear and the atoms would adiabatically follow the trap minimum during the bubble formation. 

The control of the optical bubble parameters will allow us to study the shell dynamics measuring, for instance, the frequencies of the collective modes probing the presence or the absence of an inner boundary \cite{Lannert2007,Tononi2022,Wolf2022,huang2025PRRes,Sun2018}. Changing the radius of the bubble, and thus the curvature of the surface, can impact diverse physical phenomena such as Efimov physics \cite{Naidon_2017,DIncao_2018} or the dynamics of the vortices \cite{Padavic2020,Bereta2021,Caracanhas2022}. 
 
For rubidium, the limited lifetime due to the large natural linewidth is a significant drawback of our proposal and is expected to restrain the studies of these physical phenomena.  

Our model is general and valid for all three-level systems and could be adapted to other atomic species, in particular the alkaline-earth atoms for which the narrow transition linewidth could increase the lifetime in the all-optical bubble shaped trap. Laser cooling techniques to produce Strontium  Bose-Einstein condensates\cite{Stellmer2013,Chen2022} 
have demonstrated precise engineering of strong light shifts in the excited state of a narrow transition linewidth without significant heating. These results make us confident that future investigations will validate the feasibility of this method for alkaline-earth atoms.
 
\renewcommand{\thesection}{APPENDIX \Alph{section}}
\setcounter{section}{0}
\section{Doubly dressed state potential derivation}
\label{Meth:DDSderivation}

Here, we calculate the doubly dressed state potential in the cartesian coordinates $\{\vec{x},\vec{y},\vec{z}\}$ with $\vec{\nabla}=\frac{\partial}{\partial x}\vec{x}+\frac{\partial}{\partial y}\vec{y}+\frac{\partial}{\partial z}\vec{z}$. We focus on the potential in a single direction (1D) before being expanded in 3D considering the spherical symmetry of the bubble. Under the effect of the dressing laser 1, the potential can be derived along $\vec{x}$:
\begin{multline}
    U(x',y,z) = -\int_{-\infty}^{x'} \vec{f}_\text{dip}(x,y,z).\vec{x} dx   + \\
    \int_{-\infty}^{x'} \left[ \Pi_1(x,y,z) \frac{\partial }{\partial x} U_1(x,y,z) + \Pi_2(x,y,z) \frac{\partial}{\partial x} U_2(x,y,z) \right] dx
    .
    \label{eq:general_integral_DDS_1d_1}
\end{multline}
The conservation of population $\Pi_1 + \Pi_2 = 1$ and Eq. (\ref{eq:general_integral_DDS_1d_1}) leads to:
\begin{multline}
    U(x',y,z) = \int_{-\infty}^{x'} \frac{\partial }{\partial x} U_1(x,y,z) dx     +   \\
    \int_{-\infty}^{x'}   \Pi_2(x,y,z)\frac{\partial }{\partial x} \left[ U_2(x,y,z)-U_1(x,y,z) \right] dx
    .
    \label{eq:general_integral_DDS_1d_2}
\end{multline}
The laser 2 drives the transition between the two dressed states, with the excited state population given by the optical Bloch equations with a light-shift-dependent detuning $\hbar(\Delta(x,y,z)-\Delta_\text{0})=U_1(x,y,z)-U_2(x,y,z)$ and population $\Pi_2(\Delta(x,y,z))=\frac{s}{1+s+4(\Delta(x,y,z)/\Gamma)^2}$ where $\Gamma$ is the spontaneous emission rate of the excited state $\ket{2}$ and $s$ the saturation parameter. 
Substituting for $\Pi_2$ and integrating with respect to the detuning
the second term in Eq. \eqref{eq:general_integral_DDS_1d_2} simplifies to: 
\begin{multline}
 - \hbar \int_{-\infty}^{x'}   \Pi_2(\Delta)\frac{\partial }{\partial x} \Delta (x,y,z) dx = - \hbar \int_{ \Delta(-\infty,y,z)}^{ \Delta(x',y,z)}   \Pi_2(u) du = \\
 -\frac{\hbar\Gamma s}{4\sqrt{1+s}} \left[ \atan\left( \frac{2u}{\Gamma \sqrt{1+s}} \right) \right]_{u=\Delta(-\infty,y,z)}^{u=\Delta(x',y,z)}
    \label{eq:general_integral_DDS_1d_3}
\end{multline}
where $\Delta(-\infty,y,z)=\Delta_{0}$ as both potentials vanish at infinity.
The total doubly-dressed-state potential in Eq. \eqref{eq:general_integral_DDS_1d_2} is finally analytically given by:
\begin{multline}
    U(x,y,z) = U_\text{1}(x,y,z) \\
    - \frac{\hbar \Gamma s}{4\sqrt{1+s}} \left(  \atan\left[\frac{2\Delta(x,y,z)}{\Gamma \sqrt{1+s}}\right] -
    \atan\left[\frac{2\Delta_{0}}{\Gamma \sqrt{1+s}}\right]  \right)
    .
    \label{eq:general_integral_DDS_1d_Methods}
\end{multline}
We checked that this 1D calculation is also valid in 3D, corresponding to Eq. \eqref{eq:general_integral_DDS_1d_4} in the main text. In the following we consider a potential $U(r)$ with a spherical symmetry, $r$ being the radial direction.

We calculate the first derivative of Eq. \eqref{eq:general_integral_DDS_1d_4} to evaluate the radius of the bubble $r_\text{bubble}$:
\begin{multline}
    \frac{dU(r)}{dr} = -\frac{2\Tilde{U}_1}{R_0^2} r \\
      + \frac{1}{2}\frac{s}{{1+s}} 2 r \left( \frac{\Tilde{U}_1}{R_0^2} - \frac{\Tilde{U}_2}{R_0^2} \right)  \frac{1}{1+\left[\frac{2(U_1(r)-U_2(r)+\hbar\Delta_{0})}{\sqrt{1+s}\hbar\Gamma}\right]^2}
    \label{sec:derivative_potential_DDS_3D}
\end{multline}

The extrema of the double dressed state potential are found for $\frac{dU(r)}{dr} = 0$ leading to $r_0=0$ and
\begin{equation}
    r_{\pm}^2 = (1-\eta)R_0^2 \pm \frac{\hbar \Gamma \sqrt{1+s} R_0^2}{2(\Tilde{U}_2-\Tilde{U}_1)} \sqrt{\frac{s}{1+s}\frac{\Tilde{U}_2-\Tilde{U}_1}{2|\Tilde{U}_1|}-1}
    \label{sec:derivative_potential_DDS_3D_zero}
\end{equation}
where
$r_+$ is the bubble radius and $r_-$ the position of the inner potential barrier (see Fig. \ref{fig: generic bubble potential}). Considering the low saturation regime ($s\ll1$), and considering that $\Tilde{U}_2\gg|\Tilde{U}_1|$, we find:
\begin{equation}
\begin{aligned}
    r_\text{bubble}^2 & = r_+^2 = (1-\eta) R_0^2  + \frac{\hbar\Gamma}{2\Tilde{U}_2} R_0^2 \sqrt{\beta -1}  \\
    r_\text{barrier}^2 & = r_-^2 = (1-\eta) R_0^2  - \frac{\hbar\Gamma}{2\Tilde{U}_2} R_0^2 \sqrt{\beta -1}  \\
    \label{sec:zqfsdgf}
\end{aligned}
\end{equation}
where we defined a parameter related to the ratio of the lightshifts to simplify the expressions
\begin{equation}
    \beta = \frac{s \Tilde{U}_2}{2 |\Tilde{U}_1|}
    .
\end{equation}

The effective detuning at the location of the atoms $\Delta(r_\text{bubble}) = -\Gamma/2 \sqrt{\beta -1}$ is used to calculate the population in the excited state $\ket{2}$ inside the bubble trap
\begin{equation}
\Pi_2(r_\text{bubble})=\frac{s}{1+s+\frac{4\Delta(r_{\rm bubble})^2}{\Gamma^2}}
\label{eq:PopulationState2}
\end{equation}

which impacts directly the diffusion time 
\begin{equation}
    \tau_d = \frac{1}{\Gamma \cdot \Pi_2(r_{\text{bubble}})} = \frac{ 2 + {\Tilde{U}_2}/{|\Tilde{U}_1|} }{\Gamma}. \label{eq:lifetimeDetailed}
\end{equation}

The central barrier height  $U_0 = U(r_\text{barrier}) - U(r_\text{bubble}) $ (see Fig. \ref{fig: generic bubble potential}) is estimated from the analytical expression of the potential (Eq. \ref{eq:general_integral_DDS_1d_Methods}) and can be simplified with the approximation $\beta \gg 1$ and $s \ll 1$: 
\begin{equation}
\begin{aligned}
   U_0 &= -\frac{\abs{\Tilde{U}_1}\hbar\Gamma}{\Tilde{U}_2}\sqrt{\beta - 1 } + \frac{s \hbar\Gamma}{2 \sqrt{1+s}} \atan(\sqrt{\frac{\beta -1 }{1+s}}) \\
    &\approx \hbar \Gamma \left( -2 \sqrt{\frac{s |\Tilde{U}_1|}{2 \Tilde{U}_2}} + \frac{s \pi}{4}\right)\\
    &\approx s\hbar \Gamma \frac{\pi}{4} \left(1-\frac{4}{\pi\sqrt{\beta}} \right).
\end{aligned}
\end{equation}
%
%
Similarly, the trap depth $U_t=-U(r_\text{bubble})$ leads to the following expression using the same approximations:
\begin{equation}
\begin{aligned}
    U_t &= |\Tilde{U}_1| \Bigl(\eta + \frac{\hbar \Gamma}{2\Tilde{U}_2} \sqrt{\beta -1 }\Bigl) \\
    - \frac{s \hbar \Gamma}{4 \sqrt{1+s}}&\left[\atan(-\sqrt{\frac{\beta -1 }{1+s}}) + \atan(\frac{2 \eta (\Tilde{U}_2-\Tilde{U}_1)}{\sqrt{1+s} \hbar \Gamma }) \right]  \\ 
   &\approx  \eta |\Tilde{U}_1| - \hbar \Gamma \sqrt{\frac{s |\Tilde{U}_1|}{2 \Tilde{U}_2}}\\
   &\approx  \eta |\Tilde{U}_1| - \frac{s\hbar\Gamma}{2\sqrt{\beta}}.
\end{aligned}
\end{equation}

\bigskip

\section{Calculation of the light shifts}
\label{App:LightShifts}

We present here the model used to calculate the light shifts in our numerical method to determine the double dressed states potential. We define the electric field oscillating at angular frequency $\omega$ with its complex conjugate:
\begin{equation}
\mathbf{E}=E_0 \exp{i\omega t}\mathbf{\epsilon} ~+~  \text{c.c.}
          ,
\label{eq:ElectricField}     
\end{equation}
where $\mathbf{\epsilon}$ is the unit polarization vector. The interaction of the electro-magnetic field with an atom of dipole moment $\mathbf{d}$ leads to the AC Stark shift Hamiltonian $H_\text{Stark}= -\mathbf{d}.\mathbf{E}$ which can be written in the basis of the fine structure basis\cite{STECK}:

\begin{multline}
H_\text{\rm Stark} = -\frac{1}{4}E_0^2  \Biggl( \alpha^{(0)}_{nJ} - i\alpha^{(1)}_{nJ} \frac{( \mathbf{\epsilon^{*}} \wedge \mathbf{\epsilon} ).\mathbf{\hat{J}} }{2J}  \\ +\alpha^{(2)}_{nJ} \frac{3( (\mathbf{\epsilon^{*}}.\mathbf{\hat{J}})( \mathbf{\epsilon}.\mathbf{\hat{J}})  +  (\mathbf{\epsilon}.\mathbf{\hat{J}})( \mathbf{\epsilon^{*}}.\mathbf{\hat{J}})) - 2.\mathbf{\hat{J}}^{2}}{2J(2J-1)} \Biggl)
\label{eq:StarkHamiltonian}
\end{multline}
where $\mathbf{\hat{J}}$ is the fine structure angular momentum and $\alpha^{(0)}_{nJ}, \alpha^{(1)}_{nJ}, \alpha^{(2)}_{nJ}$ are the scalar, vectorial and tensor polarizabilities of the state $\ket{n,J}$ respectively. 

In the case of a linear polarization, $\mathbf{\epsilon}$ is real and the term $\mathbf{\epsilon^{*}} \wedge \mathbf{\epsilon}$ vanishes. Thus the Stark Hamiltonian can be simplified: 
\begin{equation}
H_{\rm Stark} = -\frac{1}{4}E_0^2 ( \alpha^{(0)}_{nJ}  + \alpha^{(2)}_{nJ} \hat{Q})
,
\end{equation}
where $\hat{Q} = \frac{3 \hat{J_z}^2 - \mathbf{\hat{J}}^2}{J(2J-1)}$.

The polarisabilites can be written\cite{Arora2007}:
\begin{equation}
\alpha^{(0)}_{nJ}(\omega)=\frac{2}{3(2J+1)}\frac{1}{\hbar}\sum_{n'J'}\frac{\abs{\bra{n',J'}|\hat{d}|\ket{n,J}}^2\left(\omega_{nJ}-\omega_{n'J'}\right)}{\left(\omega_{nJ}-\omega_{n'J'}\right)^2-\omega^2}   
\end{equation}
with $\hbar \omega_{nJ}$ as the energy of the state $\ket{n,J}$ and
\begin{multline}
\alpha^{(2)}_{nJ}(\omega)= -4C\frac{1}{\hbar} \times \\ \sum_{n'J'}(-1)^{J-J'+1}\left\{\begin{matrix}J & 1 & J'\\1 & J & 2\end{matrix}\right\}\frac{\abs{\bra{n',J'}|\hat{d}|\ket{n,J}}^2\left(\omega_{nJ}-\omega_{n'J'}\right)}{\left(\omega_{nJ}-\omega_{n'J'}\right)^2-\omega^2}
\end{multline}
with
\begin{equation}
C=\sqrt{\frac{40J(2J-1)}{3(J+1)(2J+1)(2J+3)}} 
\,.
\end{equation}

The hyperfine Hamiltonian is the interaction between the nuclear $\mathbf{\hat{I}}$ and total electron $\mathbf{\hat{J}}$ angular momenta which obey the usual angular momentum properties.The sum of
these angular momenta gives the total angular momentum $F = I + J$. The hyperfine Hamiltonian, including
magnetic-dipole and electric-quadrupole terms, is given by:
\begin{equation}
    H_\text{hfs} = A_\text{hfs} \frac{\hat{\mathbf{I}}.\hat{\mathbf{J}}}{\hbar^2} + B_\text{hfs} \frac{ \frac{3}{\hbar^2} (\hat{\mathbf{I}}.\hat{\mathbf{J}})^2 + \frac{3}{2 \hbar} \hat{\mathbf{I}}.\hat{\mathbf{J}} - \hat{\mathbf{I}}^2.\hat{\mathbf{J}}^2 }{2I(2I-1)J(2J-1)}\label{eq:HyperfineHamiltonian}
\end{equation}
where $A_\text{hfs}$ is the magnetic dipole hyperfine constant and $B_\text{hfs}$ the electric-quadrupole hyperfine constant \cite{STECK}.

For large light shifts compared to the hyperfine splittings, hyperfine states $\ket{F}$ mix. Eigenstates will be a superposition of hyperfine states that tend to the fine structure eigenstates $|m_j|$ in the strong-field limit. Note that the strong-field regime is analogous to the Paschen-Back regime for magnetic fields and requires us to diagonalize the full Hamiltonian $H_{\rm Stark}+H_{\rm hfs}$.

\section{Energy spectrum without interaction}
\label{App:states_without_interaction}

In the non interacting regime, we calculate the eigenvalues of the Hamiltonian composed of the kinetic and potential energies in the radial direction of the bubble-shaped trap. We compare three different potentials in order to find an approximate expression of the energy spectrum. 

The potential energy is defined by the full doubly-dressed state potential (DDS):
\begin{multline}
    U_\text{DDS}(r) = U_\text{1}(r) \\
    - \frac{\hbar \Gamma s}{4\sqrt{1+s}} \left(  \atan\left[\frac{2\Delta(r)}{\Gamma \sqrt{1+s}}\right] -
    \atan\left[\frac{2\Delta_{0}}{\Gamma \sqrt{1+s}}\right]  \right)
    .
    \label{eq:general_integral_DDS_1d_Methods_2}
\end{multline}
The Schr\"{o}dinger equation with this potential does not have simple solutions and the spectrum has to be found numerically.

A good approximation of the DDS potential
is the hard-wall limit, where the "atan" terms can be simplified as a potential barrier of amplitude $U_0$. In this case, we define the hard-wall harmonic oscillator (HWHO) potential as:
\begin{equation}
    U_\text{HWHO}(r) = \left\{
        \begin{array}{ll}
            U_\text{1}(r) & \mbox{if } r > r_\text{barrier} \\
            U_0 & \mbox{if } r\leq r_\text{barrier}
        \end{array}
    \right.
    \label{eq:HWHO_potential}
\end{equation}
For $r_\text{barrier}=0$, the problem is equivalent to the half harmonic oscillator whose solutions are the same as the full harmonic oscillator except that the Hermite functions cancel $H_n(0)=0$, which is the case for odd $n$. For a non-zero value of $r_\text{barrier}$, the solutions verify $H_n(r_\text{barrier})=0$ and are not simple generally..

A second approximated potential is the half-wedge potential:
\begin{equation}
    U_\text{wedge}(r) = \lambda_r |r|
    \label{eq:wedge_potential}
\end{equation}
where the potential varies linearly as $\lambda_r$. 

For the potential given by Eq. (\ref{eq:wedge_potential}), the eigenstates are given by Airy functions. Inspired by Gallas et al. \cite{Gallas1995ER}, we can write the eigenvalues for large arguments of the Airy function as: 
\begin{equation}
    E_{n,\text{wedge}} = \frac{1}{8} \left( \frac{\hbar^2\lambda_r^2}{m} \right)^{1/3} \left[6\pi\left( 2n+1 \right)\right]^{2/3}
    \label{eq:wedge_eigenvalues}
\end{equation}
where $n$ is in $\{0,1,2,...\}$.

Finally, we can find approximated analytical eigenvalues of the bubble potential by using the results of the half-wedge (HW) potential, with a linearised potential such that: 
\begin{equation}
    U_\text{half wedge}(r) = \left\{
        \begin{array}{ll}
          \frac{dU_1(r)}{dr}|_{r_\text{barrier}} |r-r_\text{barrier}| & \mbox{if } r > r_\text{barrier} \\
            U_0 & \mbox{if } r\leq r_\text{barrier}
        \end{array}
    \right.
    \label{eq:HW_potential}
\end{equation}
where $\frac{dU_1(r)}{dr}|_{r_\text{barrier}}=-2\frac{\tilde{U}_1}{R_0^2}r_\text{barrier}$ that is evaluated at the position $r_\text{barrier}$.

The eigenvalues of the half-wedge potential given by Eq. (\ref{eq:HW_potential}) require that the wave-functions vanish at the position of the barrier which corresponds to selecting all the odd solutions of the full wedge potential. Thus the eigenvalues of the half-wedge are given by Eq. (\ref{eq:wedge_eigenvalues}) for odd $n$, resulting in the eigenvalues for the half-wedge:
\begin{equation}
    E_{n,\text{HW}} = \frac{1}{8} \left( \frac{\hbar^2\lambda_r^2}{m} \right)^{1/3} \left[6\pi\left( 4n+3 \right)\right]^{2/3}
    \label{eq:half_wedge_eigenvalues}
\end{equation}
where $n$ is in $\{0,1,2,...\}$.


To determine the dimensionality, the relevant energy scale in the radial direction of the bubble is the energy difference between the first and second excited state in the half-wedge limit:
\begin{equation}
\begin{aligned}
    \frac{\Delta E_{1,0}}{\hbar}&\equiv \frac{E_{1,\text{HW}} - E_{0,\text{HW}}}{\hbar} \\
    & = \frac{1}{8} \left( \frac{4  {\tilde{U}_1^2}r^2_\text{barrier} }{m \hbar R_0^4} \right)^{1/3} \left(6\pi\right)^{2/3} \left(  7 ^{2/3} - 3^{2/3} \right) \\
    & \approx 2.22 \left( \frac{  {\tilde{U}_1^2}r^2_\text{barrier} }{m \hbar R_0^4} \right)^{1/3}
    \label{eq:HW_trap_frequency}
\end{aligned}
\end{equation}  
which is the expression given in the main text.

\begin{figure}[!htb]
    \centering
    \includegraphics[width=1\linewidth]{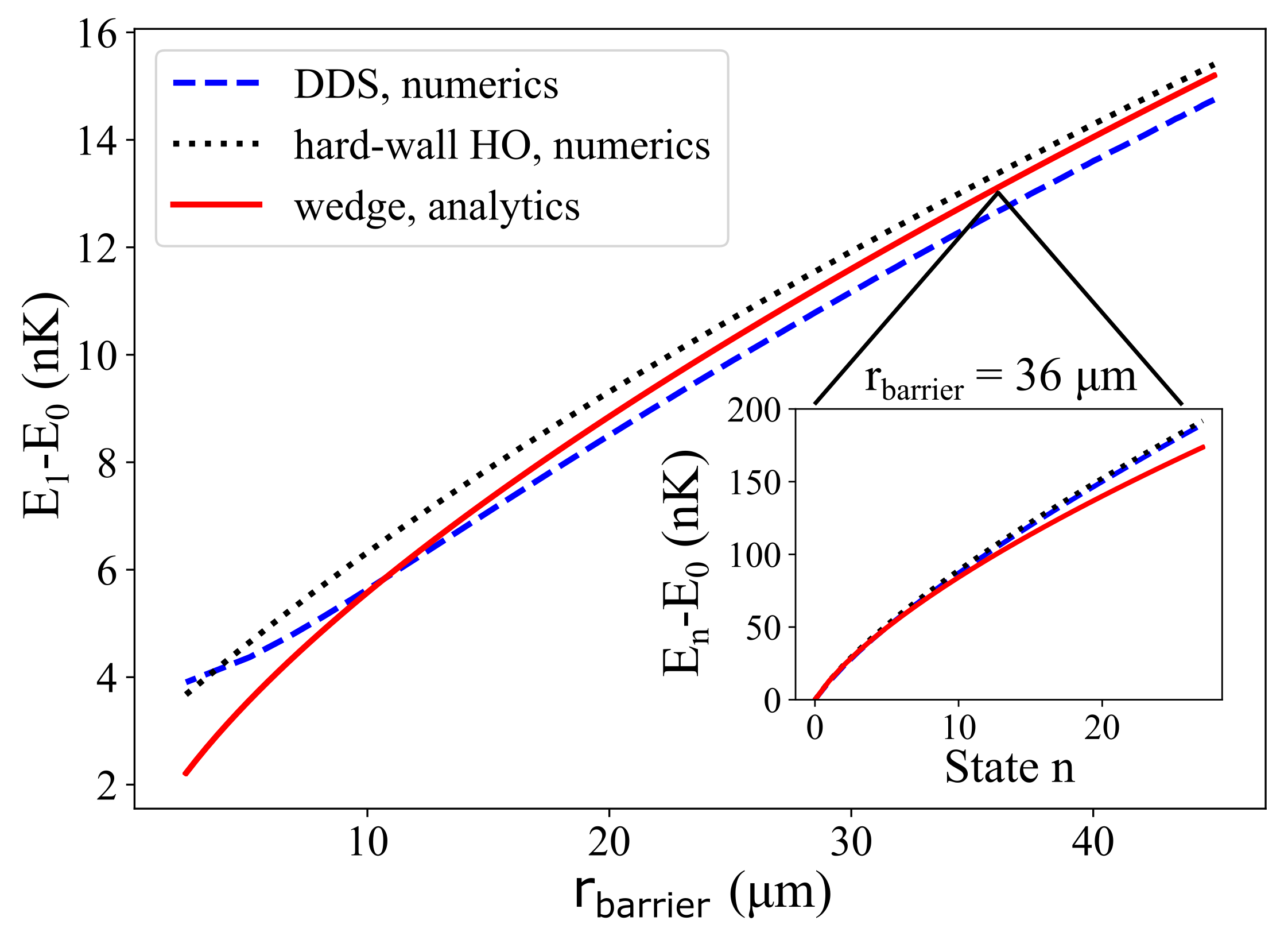}
    \caption{Comparison of the energy difference for the first two eigenvalues versus the position of the barrier, calculated numerically for the hard-wall harmonic potential (HO) and the doubly-dressed state potential (DDS), and analytically for the half-wedge potential using Eq. (\ref{eq:half_wedge_eigenvalues}). The inset is an example of the eigenenergies for a bubble radius of 36 {\textmu}m, with 27 bound states.}
    \label{fig:comparison_eigenstates}
\end{figure}

In Fig. \ref{fig:comparison_eigenstates}, we compare the eigenstates calculated numerically for the three above potentials. As shown in the inset, the HWHO numerical model (black dotted line) describes very well the energy spectrum of the full DDS potential (blue dashed line). The half-wedge is in good approximation close to the other for small $n$ and deviates at larger $n$ due to the linear approximation. However, for small $n$, all the models are in good agreement. The energy difference between the first two states are consistent as well, with stronger deviations for small $r_\text{barrier}$, also due to the linear approximation which starts to fail for small $r$.

The energy of the different eigenstates with the different angular momentum number $l$  complete the energy spectrum:
\begin{equation}
\epsilon_l=\frac{\hbar^2 l(l+1)}{2 m r_{\rm bubble}^2}
.
\label{eq:EnergyStateAngularMomentum}    
\end{equation}
For $l=1$, $r_{\rm bubble}=35$ \textmu{m}, we get $\epsilon_1=$2 pK, which is much smaller than the energy difference between two radial states, for the range of bubble radii considered in this paper (see Fig. \ref{fig:comparison_eigenstates}).

The critical temperature for condensation of non interacting Rb Bosons is the solution of the equation\cite{Tononi2019}:
\begin{equation}
k_B T_C=2\pi\frac{\hbar^2 n}{m}\left[\frac{\hbar^2}{mr_{\rm bubble}^2 k_B T_C}-\ln{\left[e^{\hbar^2/(mr_{\rm bubble}^2k_B T_C)}-1\right]}\right]^{-1}
\label{eq:CriticalBECNonInteracting}
\end{equation}
which can be solved numerically with $n=N/(4\pi r_{\rm bubble}^2)$ the 2D density. With $10^5$ atoms and $r_{\rm bubble}=35$ \textmu{m}, $T_c$ is 25 nK \cite{TONONI2024}.  We expect interactions in the mean field regime won't significantly change the situation at very low temperature.

\section{Energies and Wave-Functions in the interacting regime}
\label{App:Numerical model}

We developed a numerical model solving the Gross-Pitaevski Equation (GPE) to calculate the wave-functions in the ground state at zero temperature and determine the evolution of the kinetic, potential and interaction energies when we vary the parameters of the bubble shaped trap (see Figure \ref{fig:EnergiesversusRadius} and \ref{fig : variation laser power 2 1D 2 laser gaussian 1529 and 770} in section \ref{sec:EnergyScale}).

To compute the wave-functions and the energies of the different levels in the bubble shaped trap, we consider a perfectly spherical symmetry and use the imaginary time propagation. Taking into account the atomic interactions, we write the 3D Gross Pitaevski Equation  (GPE) out in spherical coordinates:
\begin{equation}
i\hbar \frac{\partial \tilde{\Psi}}{\partial t} = \left( -\frac{\hbar^2}{2m} \nabla^2 + V(\mathbf{r}) + gN_\text{at} |\tilde{\Psi}|^2 \right) \tilde{\Psi}
\end{equation}
where the Laplacian in spherical coordinates is $\nabla^2 = \frac{1}{r^2} \pdv{}{r} \left( r^2 \pdv{}{r} \right)
- \frac{\vec{L}^2}{r^2}$ with $\vec{L}$ the orbital angular momentum operator. 

Due to the spherical symmetry, we look for solutions that are separable as \cite{GriffithsBook2018,Bereta2019} $\tilde{\Psi}(r,\theta,\phi,t)=\Psi_{l}(r,t)Y_{lm}(\theta,\phi)$ where the spherical harmonics verify the eigen-equation identity $\hat{L}^2 Y_{lm}(\theta, \phi) = \ell(\ell + 1) \hbar^2 Y_{lm}(\theta, \phi)$. In addition, we focus on the lowest-energy condensate $l=0$ for which $Y_{00}=1/\sqrt{4\pi}$, leading to the radial evolution:  
\begin{equation}
    i \hbar \frac{\partial \Psi }{\partial t } = \frac{- \hbar ^2}{2m} \frac{1}{r^2} \frac{\partial}{\partial r } \left(r^2 \frac{\partial \Psi}{\partial r }\right) + V(r) \Psi + \frac{gN_\text{at}}{4\pi} |\Psi|^2 \Psi \label{eq:3DGPE}
    .
\end{equation}
If $l\neq0$, the potential can be corrected with an extra term due to the centrifugal potential $\hbar^2l(l+1)/(2mr^2)$ that is equal to zero for $l=0$.

With the renormalization condition: 

\begin{equation}
\iiint  |\tilde{\Psi}(r,\theta,\phi)|^2 r^2\sin(\theta) d\theta d\phi dr = \int_0^{\infty} r^2 |{\Psi(r)}|^2 dr = 1
\label{eq:Renormalization}
\end{equation}
leading us to consider the 1D wave-function $ {\Psi}^{(1D)}(r,t) = r {\Psi}(r,t)$ with $\int_0^{\infty} |\Psi^{(1D)}(r,t)|^2 dr = 1$.

We derive the following 1D Gross Pitaevski equation: 
\begin{equation}
    i \hbar \frac{\partial \Psi^{(1D)} }{\partial t } = \frac{- \hbar ^2}{2m} \frac{\partial^2 \Psi^{(1D)}}{\partial^2 r^2 } + V(r) \Psi^{(1D)} + \frac{g N_\text{at}}{4 \pi r^2} |\Psi^{(1D)}|^2 \Psi ^{(1D)}
    \label{eq:1DGPE}
\end{equation}
\begin{equation}
 \approx \frac{- \hbar ^2}{2m} \frac{\partial^2 \Psi^{(1D)}}{\partial^2 r^2 } + V(r) \Psi^{(1D)} + \frac{g N_\text{at}}{4 \pi r_{\rm bubble}^2} |\Psi^{(1D)}|^2 \Psi ^{(1D)}
 \label{eq:1DGPEbis}
\end{equation}
with the approximation related to the localization of the wave-function on the surface of the bubble.    
The ground state of our bubble-shape potential is found using the imaginary time method.


We calculate independently the different energies terms using the calculated wave-function, leading to the kinetic energy:
\begin{equation}
E_{kin}= \iiint \frac{\hbar^2}{2m}r^2|\nabla \tilde{\Psi}(r,\theta,\phi)|^2 d\theta d\phi d r
\label{eq:KineticEnergy}    
\end{equation}

the potential energy:
\begin{equation}
E_p=\iiint V(r)r^2|\tilde{\Psi}(r,\theta,\phi)|^2 d\theta d\phi d r
\label{eq:PotentialEnergy}    
\end{equation}

and the interaction energy:
\begin{equation}
E_i= N_\text{at} \iiint \frac{g}{2}r^2|\tilde{\Psi}(r,\theta,\phi)|^4 d\theta d\phi  d r
    \label{eq:InteractionEnergy}
\end{equation}

The chemical potential in the case of the well-wedge potential, as well as the hard-wall harmonic oscillator (HWHO), is difficult to derive directly. To simplify the calculation of an analytical expression, we consider an asymmetric harmonic (AHO) trap with two different trap frequencies $\omega_{-} \ll \omega_{+}$:
\begin{equation}
    U_\text{AHO}(r) = \left\{
        \begin{array}{ll}
          \frac{1}{2} m \omega_-^2 (r-r_{\rm bubble})^2 = b_- (r-r_{\rm bubble})^2 & \mbox{if } r \leq r_{\rm bubble} \\
            \frac{1}{2} m \omega_+^2 (r-r_{\rm bubble})^2 = b_+ (r-r_{\rm bubble})^2 & \mbox{if } r  \geq r_{\rm bubble}
        \end{array}
    \right.
    \label{eq : HHO_potential}
\end{equation}
with $b_{\pm}=1/2 m \omega_{\pm}^2$. In the Thomas Fermi regime the radial atomic density depends on the chemical potential $\mu$: 
\begin{equation} \label{eq : n(r) TF}
    n(r) = \max \left[ 0 , \frac{\mu - U_\text{AHO}(r)}{g_{3D}} \right]  
\end{equation} 
with the interaction parameter $g_{3D} = 4 \pi \hbar^2 a/m$. We integrate the atomic density (Eq. \ref{eq : n(r) TF}) to get the total atom number: 
\begin{equation}
    \begin{aligned}
        & N_\text{at} =  \int_{0}^{+\infty} 4 \pi r^2 n(r) dr = \\
        & \frac{4 \pi}{g_{3D}} [ \int_{r_{\rm bubble}-\sqrt{\frac{\mu}{b_-}}}^{r_{\rm bubble}} r^2 (\mu-b_- (r-r_{\rm bubble})^2) dr + \\
        & \int_{r_{\rm bubble}}^{r_{\rm bubble}+\sqrt{\frac{\mu}{b_+}}} r^2 (\mu-b_+ (r-r_{\rm bubble})^2) dr ] \\
    \end{aligned}
\end{equation}

The total atom number can be expressed as a function of the reduced parameter $\epsilon_{\pm} = \frac{\mu}{b_{\pm} r_{\rm bubble}^2}$:  
\begin{equation}
    \begin{aligned}
        & N_\text{at} = \frac{4\pi}{g_{3D}} \frac{r_{\rm bubble}^3 \mu}{30} ( 20 \sqrt{\epsilon_-} + 4\epsilon_-\sqrt{\epsilon_-} - 15\epsilon_- \\
        & + 20 \sqrt{\epsilon_+} + 4\epsilon_+\sqrt{\epsilon_+} + 15\epsilon_+) \\
    \end{aligned}
\end{equation}

We consider here the situation of the thin bubble shaped trap corresponding to $ \epsilon_{\pm} \ll 1 $: 
\begin{equation}
        N_\text{at} = \frac{2\sqrt{2} \sqrt{m} r_{\rm bubble}^2 \mu^{3/2}}{3a \hbar^2} \frac{\omega_+ + \omega_-}{\omega_+ \omega_-}
\end{equation}

In the limit of the hard wall $\omega_{-} \rightarrow \infty$ and relating the trap frequency to the energy difference of the two first states $ \hbar\omega_+ = \Delta E_{1,0}$:
\begin{equation}
        \mu = \left ( \frac{3 N_\text{at} a \hbar \Delta E_{1,0}}{2\sqrt{2} \sqrt{m} r_{\rm bubble}^2} \right ) ^{2/3}
\end{equation}

We checked that this analytic formula is consistent with our numerical model of the DDS potential. 
This result leads to an analytical criterion on the atom number to be in the quasi-2D regime $\mu < \Delta E_{1,0}$:
\begin{equation}
     N_\text{at} < \left( \frac{8 m^2 |\tilde{U}_1|^2 (1-\eta)^7 R_0^{10}}{a^6 \hbar^4}    \right) ^{1/6}
\end{equation}
using $r_{\rm barrier}^2\approx r_{\rm bubble}^2 \approx (1-\eta)R_0^2$. For the parameters presented in section \ref{sec:EnergyScale}, the quasi 2D regime is fulfilled for atom numbers as large as $N_\text{at}<2\cdot 10^5$.

\section*{Acknowlegdment}
The authors gratefully acknowledge discussions with Philippe Bouyer. This work is supported by the French national agency CNES (Centre National d’Etudes Spatiales), the European Space Agency (ESA), the Investments for the Future Programme IdEx Bordeaux-LAPHIA (ANR-10-IDEX-03-02), ANR contracts (JCJC ANR-18-CE47-0001-01 and QUANTERA21 ANR-22-QUA2-0003) and the Quantum Matter Bordeaux.  For financial support, C.Metayer thanks CNES and ESA. For financial support, R. Veyron thanks "NextGenerationEU/PRTR" (Grant FJC2021-047840-I) and "Severo Ochoa" Center of Excellence CEX2019-000910-S; Generalitat de Catalunya through the CERCA program, DURSI grant No. 2021 SGR 01453 and QSENSE (GOV/51/2022). Fundaci\'{o} Privada Cellex; Fundaci\'{o} Mir-Puig.

\section*{Author Declarations}

\subsection*{Conflict of interest}
The authors have no conflicts to disclose.

\subsection*{Author Contributions}


\textbf{R. Veyron}: Formal Analysis (equal), Software (equal), Investigation (equal), Writing-Original draft Preparation (equal), Writing-Review and Editing (equal).
\textbf{C. Métayer}: Formal Analysis (equal), Software (equal), Investigation (equal), Writing-Original draft Preparation (equal), Writing-Review and Editing (equal).
\textbf{J.B. Gerent}: Writing-Original draft Preparation (supporting), Writing-Review and Editing (equal).
\textbf{R. Huang}: Writing-Original draft Preparation (supporting), Writing-Review and Editing (supporting).
\textbf{E. Beraud}: Writing-Original draft Preparation (supporting), Writing-Review and Editing (supporting).
\textbf{B.M. Garraway}: Writing-Original draft Preparation (supporting), Writing-Review and Editing (equal).
\textbf{S. Bernon}: Conceptualization (equal), Methodology (equal), Writing-Original draft Preparation (equal), Writing-Review and Editing (equal).
\textbf{B. Battelier} : Funding Acquisition (lead), Validation (lead), Supervision (lead), Conceptualization (equal), Methodology (equal), Writing-Original draft Preparation (equal), Writing-Review and Editing (equal).

\section*{Data availability}

Data sharing is not applicable to this article as no new data were created or analyzed in this study.

\bibliographystyle{unsrt}
\bibliography{bibliographie}

\end{document}